\documentclass[trackchanges,twocolumn]{aastex701}

\usepackage{amsmath}

\begin{document}

\title{Minijets and Broken Stationarity in a Blazar : Novel Insights into the Origin of $\gamma$-ray Variability in CTA 102}

\author[orcid=0000-0003-1101-8436]{Agniva Roychowdhury}
\affiliation{National Centre for Radio Astrophysics - Tata Institute of Fundamental Research, Pune University Campus, Ganeshkhind, Pune 411007, MH, India}
\email[show]{agniva.physics@gmail.com}  

\begin{abstract}

High-energy blazar light curves, in X-rays and beyond, have historically preferred a log-normal flux distribution, signifying multiplicative processes either in the jet itself or due to connection(s) with accretion. Here we present 18 year archival Fermi-LAT light curves (0.1-100 GeV) of the flat spectrum radio quasar (FSRQ) CTA 102 from August 2008 to November 2025, which underwent a huge flare in 2017, with a $\sim$ factor of 100 jump in $\gamma$-ray flux, along with similar flaring in X-rays. Our statistical analyses confirm that neither the pre nor the post-flare total GeV light curves follow a strictly log-normal distribution. Instead, we observe a statistically significant reduction in skewness from the pre to the post-flare light curves, which implies the blazar transitioned from an energetic state with frequent flaring to a more plateaued state with occasional flaring. We further find that this state transition can be explained through magnetic relaxation, where many reconnection events caused the 2017 flare, after which the magnetic field was ordered and its energy reached a minimum. To explain this further, we use a Monte Carlo simulation of a modified minijets-in-a-jet model where GeV flares are produced only when a maximum number of minijets move toward the broad line region and towards the line of sight, in the context of an external Compton model. The flux distributions (both observed and simulated) could be fit by a modified log-normal power-law distribution, implying our minijets model can reproduce the GeV flares in CTA 102 as well as their flux distributions.

\end{abstract}



\section{Introduction} 

Blazars are a class of radio-loud active galactic nuclei (AGN) with a relativistic jet almost pointed close to our line of sight \citep[e.g.,][]{fos98,blandford2019}. A typical feature of blazars is their extreme variability from radio to very high energy (VHE) $\gamma$-rays, primarily driven by compact emitting regions, very high Doppler factors and a fast rate of energy injection \citep[e.g.,][]{raiteri25}. Studies of blazar variability across different wavebands can probe the physical conditions in the emitting region and constrain blazar physical parameters including the magnetic field.

In recent times, multiple studies have revealed an otherwise peculiar nature of high-energy blazar variability. The flux distribution, instead of following a simple Gaussian, follows a log-normal distribution, where the flux logarithm follows a Gaussian \citep[e.g.,][]{sinha18, bhatta20}. If independent random emitting regions were contributing to the variability, the former would have been expected. Proof of the latter confirms the presence of complex multiplicative cascade processes, where multiple emitting regions interact with and affect each other. A popular review by \cite{mitzen04} discusses the fundamental generative models that can contribute to distributions that arise in various fields of academics. In particular, a multiplicative process where $X_i=F_iX_{i-1}$ (where $F$ is a random variable) can be written out as $\log X_i=\log X_0+\sum_1^i\log F_k$, that leads to a normal distribution in the series $\log X$ using the Central Limit Theorem for $\sum_1^i \ln F_k$. In our case, simply put, the emissions from various regions pile up exponentially and produce the observed light curve. Earlier thought to be akin to accretion disk variations (where propagating accretion rate fluctuations travel from the outer to the inner disk, e.g., \citealt{lyub97}), recent studies \citep{biteau12} have shown that randomly oriented minijets (with random Doppler factors) inside a single jet and/or fluctuations of energy injection in the electron energy distribution evolution equation \citep{sinha18} can self-consistently produce flux distributions that are tailed power-laws, log-normal or Pareto-like.

This brief study focuses on a popular flat spectrum radio quasar (a sub-class of blazars itself with an optical-UV synchrotron peak and an X-ray external Compton peak) CTA 102 ($z=1.037$) \citep{schmidt65} that was seen to undergo flaring across optical to $\gamma$-rays during 2017 (e.g., \citealt{kim25}). The Fermi-LAT light curve repository contains a rich history of this source, spanning 18 years from 2008 to now, that shows smaller flares in addition to the large flare during 2017. The strong variability of this source from radio to $\gamma$-rays, with a history of extreme flaring events in the $\gamma$-rays made this an ideal choice to test variability theories in blazar jets. Further, this is one of the rare sources that contains enough flux data a decade before and after an extreme GeV flare, creating a chance for robust light curve  analysis. In this paper we focus on the GeV flux distribution of this source and conduct varied statistical analyses to test the physical origin of flaring in this blazar.

The paper is arranged is follows. Section 2 contains the primary analyses and results. Section 3 contains a brief discussion and Section 4 contains the conclusions.

\section{Analyses and Results}

\begin{figure}
    \centering
    \includegraphics[width=0.9\linewidth]{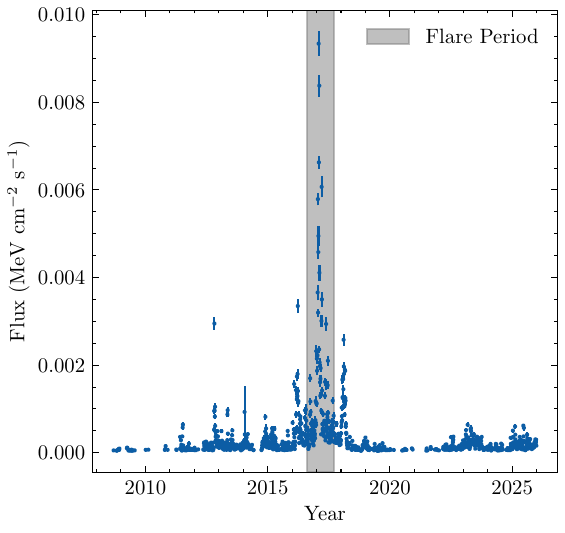}
    \caption{The 18 year Fermi-LAT light curve of CTA 102, with the gray region showing the massive flaring period.}
    \label{fig:lc}
\end{figure}

The Fermi-LAT light curve of CTA 102 was downloaded from the Fermi-LAT archive\footnote{https://fermi.gsfc.nasa.gov/ssc/data}. All points with test-statistic (TS) $<25$ were rejected. The total Fermi-LAT light curve is shown in Figure \ref{fig:lc}. The flare lasted roughly from July 2016 to August 2017 (P2, Table 1, of \citealt{kim25}), and we used the same to divide the light curve into pre-flare and post-flare regions.

\begin{figure}
    \centering
    \includegraphics[width=0.9\linewidth]{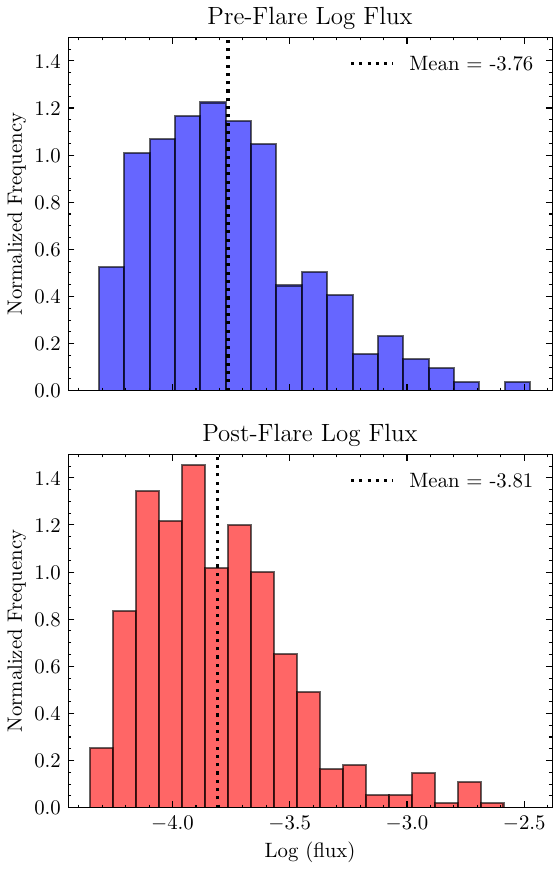}
    \caption{Figure shows the histograms of the logarithmic fluxes for the pre-flare and post-flare light curves, in blue and red, in upper and bottom panels respectively. The ``flux transfer" from the tail to the centre is evident.}
    \label{fig:hist}
\end{figure}

Figure \ref{fig:hist} shows the pre-flare and post-flare logarithmic flux distributions in blue and red respectively. Non-gaussianity is clear in both the distributions, with both having an extended tail. The pre-flare light curve, having multiple flares shows a broad tail. In contrast, the post-flare distribution shows a ``transfer" of flux from the tail towards the median, implying loss of sudden flaring events, both of which is somewhat evident from Figure \ref{fig:lc}. We ran a Kolmogorov-Smirnov (KS) test on the two distributions to find a test statistic $\simeq 0.1$ at a p-value of $p\simeq 0.06$ ($\lesssim 2\sigma$ confidence). This indicates that both the pre-flare and the post-flare light curves originate from the same distribution. However, the pre-flare flux histogram has a longer tail than the post-flare.

\begin{figure*}
    \centering
    \hbox{
    \includegraphics[width=0.5\linewidth]{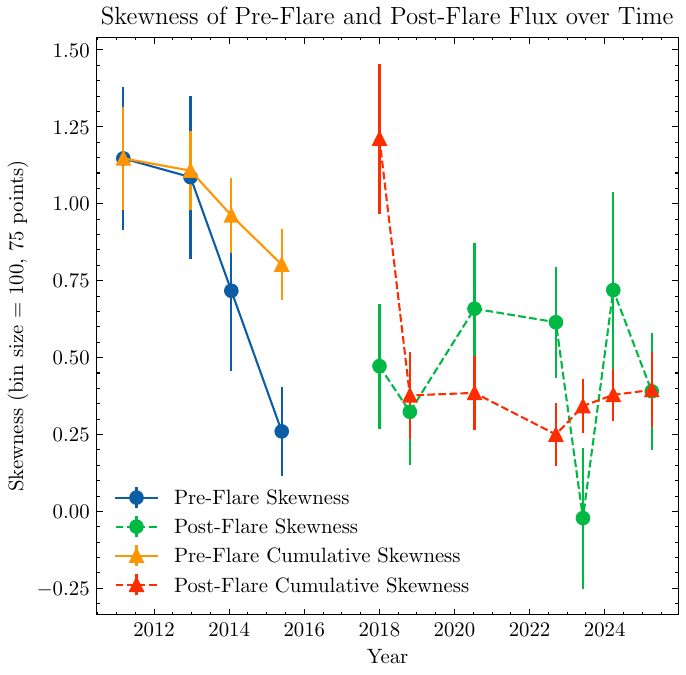}
    \includegraphics[width=0.5\linewidth]{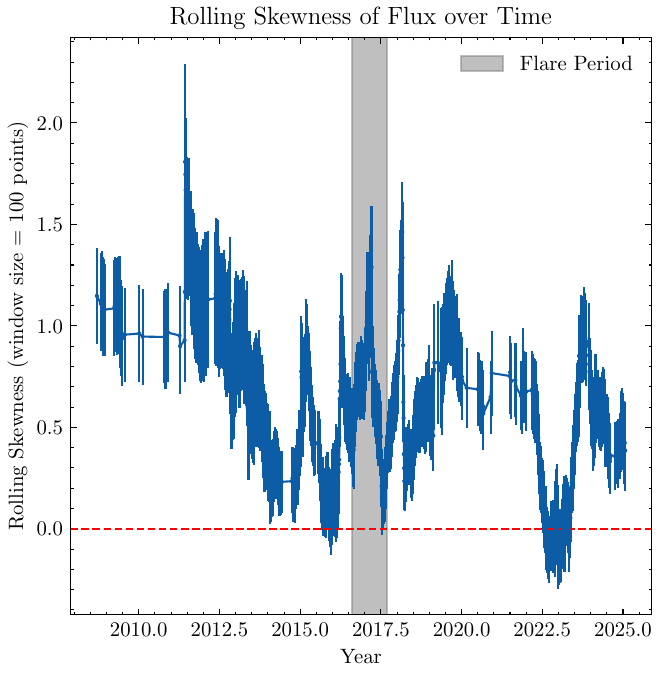}
    }
    \caption{Left panel (a) shows the ``per-1 year" bin skewness for both the pre-flare and post-flare light curves, in addition to the cumulative skewness taking more and more points into account \textit{until} the flare, from either side of the light curve. Right panel (b) shows the rolling skewness of a window of size 1 year, advanced by one point. A beginning of a transition from a high to a low skewness is visible in the post-flare state.}
    \label{fig:skew}
\end{figure*}

To further quantify the possible difference in nature of the source pre-flare and post-flare, we computed the skewness as a function of time. For each of the states, we computed the skewness of the distribution for every 100 point bin ($\sim 1$ year, to maintain statistical significance) until the flare. For the pre-flare state, it was computed from left towards right, increasing in time. For the post-flare state, we measured the skewness from the latest data point, and stopped until the flare. The measuring direction of skewness does not matter for individual bins, but we next calculated the ``cumulative" skewness, adding up all the bins in a way beginning from either side until the flare. The direction will matter in the latter. This was motivated by the objective to understand the contamination to a theoretical zero skewness by a large flare. In order to calculate the uncertainties on the skewness values, we took the entire logarithmic flux distribution, ran a boostrapping analysis with 1000 samples (random point selection with replacement), and from the resulting distribution, took the skewness value as the median $\pm\,1\sigma$. Figure \ref{fig:skew} shows both the cumulative and the per-bin skewness. The left panel (a) shows the per-bin skewness and the cumulative both. It is evident that Figure \ref{fig:skew}(a) the skewness of individual bins hovers between $\sim0.0$ and $\sim1.7$, with strong preference towards higher and lower skewness for the pre and post-flare distributions respectively. The cumulative skewness for the logarithmic fluxes, on the other hand, show distinct non-gaussianity even crossing unity for the pre-flare state.

In addition, we also computed a ``rolling" skewness, with a window size of 100 points (to maintain statistical significance) that moved by one point. The uncertainties were calculated similarly as described in the previous paragraph. The right panel (b) of Figure \ref{fig:skew} shows the rolling skewness. This tells a more complete story with a clear decrease in skewness in the short-time light curves (100 points $\sim$ 1 year) from the pre-flare to the post-flare state. This is a direct result of a transition of the distribution from having an extended tail with more flaring to having a shorter tail with more steady flux. In order to test the validity of the above observation, we ran a Mann Whitney Test \citep{mw47} on the pre and post-flare skewness distributions to confirm if the pre-flare skewness is indeed greater than the post-flare skewness. We obtained a test-statistic of $\sim10^5$ and a p-value $\sim 10^{-9}$, implying the pre-flare skewness is greater than the post-flare skewness at a very strong statistical significance.

It seems from first glance that the skewness oscillates around 0 in the post-flare state. However, we found that the post-flare skewness mean is $\sim0.5$, compared to the $\sim0.7$ mean in the pre-flare skewness. A simple t-test was used to reject this hypothesis of ``return" to Gaussianity. Instead, the post-flare flux distribution tends to stabilize toward a lower skewness, although far away from log-normality.

\begin{figure}
    \centering
    \includegraphics[width=\linewidth]{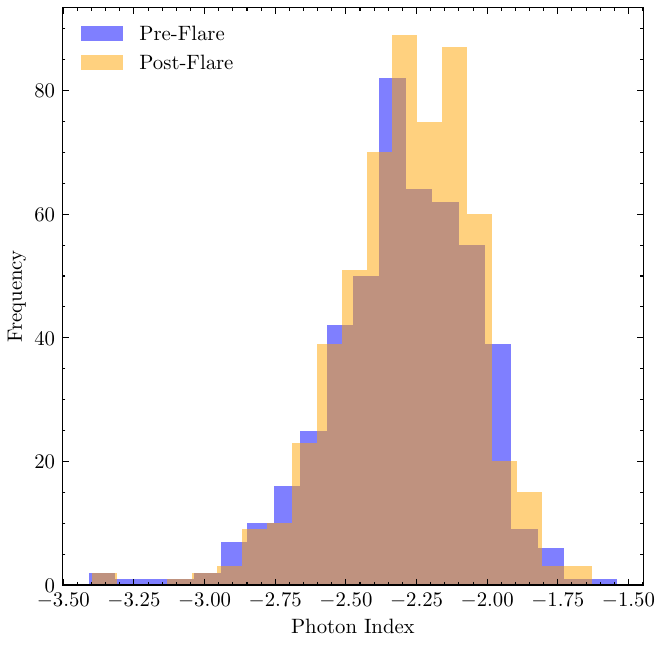}
    \caption{Figure shows the Fermi photon index histograms for the pre-flare and post-flare states.}
    \label{fig:ph_index}
\end{figure}

While the flux distribution changes through the flare, the $\gamma$-ray photon indices remain almost the same throughout. Figure \ref{fig:ph_index} shows the corresponding histograms for the pre-flare and post-flare states. They are virtually the same distribution with a KS statistic 0.04 at a $\lesssim 1\sigma$ confidence.

In the next sections we will discuss the implications of the above results in the context of blazars in particular and jet physics in general.

\section{Discussion}

Nonstationarity in a time series is generally defined as a statistically significant change in its mean and the variance over time \citep[e.g.,][]{joy20}. The distributions obtained here are not log-normal, and the pre-flare and post-flare light curves show difference in mean and especially variance for their logarithmic flux, which is higher for the pre-flare fluxes. While not log-normal, the distributions still signify a multiplicative radiative mechanism at action, where all parts of the total emitting region communicate with each other. The minijets-in-jet statistical model of \cite{biteau12} contain smaller emitting regions inside the global parsec-scale jet, which have random Doppler factors $\delta_{\rm mj}$ owing to random orientations (random angle to line of sight $\theta_{\rm mj}$ although similar speeds $\Gamma_{\rm mj}$). The total emission from such a configuration obeys a linear flux-RMS relation and produces a [long-tailed] Pareto distribution for the logarithmic flux. Our findings are different since we have skewed Gaussians at best, which are different from a Pareto function. This would imply the minijets possibly have different velocities $\Gamma_{\rm mj}$, resulting in a superposition of multiple Pareto distributions (one for each $\Gamma_{\rm mj}$). While this is an introduction of an extra variable, the scenario is very natural and there is no reason why multiple shocks/minijets in a jet would have the same Lorentz factor. In Sections \ref{sec:minijets} and \ref{sec:mlp}, we model this departure from log-normality using a modified minijets model and a Modified Log-Normal Power-Law distribution (MLP, discussed in \citealt{basu2015}), whose foundation and implications will be discussed in the following subsections.

Further, from the photon index evolution obtained in Figure \ref{fig:ph_index} it is possible that the radiative conditions have possibly not changed through the entire light curve. But due to large errors in the indices and spectral modelling being out of the scope of this paper, we cannot say for sure. However, we intend to believe the constancy of the index to first order and only invoke radiation conditions when necessary. The first possibility that would produce a constant photon index through the flare but still produce a change in the parameters of the flux distribution is changes in the geometry of the jet and/or magnetohydrodynamic (MHD) effects. A change in the external geometry of the jet across the flare is unlikely since that would only boost/de-boost the jet luminosity and add a scaling factor which would not change the skewness of the logarithmic distribution. This leaves us with its internal geometry and MHD effects.

\subsection{Minijets, the BLR, and CTA 102}
\label{sec:minijets}
The importance of this flare that occurred in $\sim 2017$ is relevant. The flare was seen in $\gamma$-rays, X-rays and the optical \citep{ma25} and was brighter by $\gtrsim7$ times the mean-level flux before the flare, while other flares in the 18 year light curve have comparable peak intensities as evident in Figure \ref{fig:lc}. The event and the ``state" change, i.e., the entire global evolution of the Fermi light curve and the blazar over the last 18 years can be fully explained by the minijets-in-jet model. We assume a very simple model of external compton (EC) scattering off broad line region (BLR) photons for the GeV emission. This is generally true for FSRQs \citep[e.g.,][]{fos98}, where EC dominates over synchrotron self-Compton (SSC). For minijets, there should be multiple GeV emission regions. If one is outside and is moving \textit{towards} the BLR, the incoming photon energy densities in the frame of the minijet will be boosted (note this is always true if the region is inside the BLR) as a function of the $\Gamma_{\rm mj}$, $\theta_{\rm mj}$. One must note that the resulting GeV flux will be boosted \textit{only if} the minijet is moving \textit{towards} us, regardless of whether it received boosted BLR photons. In the case where a maximum number of emitting regions/minijets ``line up" towards the BLR as well as are moving toward us, a massive flare can be produced. For smaller scale flares, this would imply far less number of emitting regions lined up towards the BLR and/or many minijets moving away from us. We estimate the GeV flux densities that can be obtained from this model in the next paragraph.

\begin{figure}
    \centering
    \includegraphics[width=\linewidth]{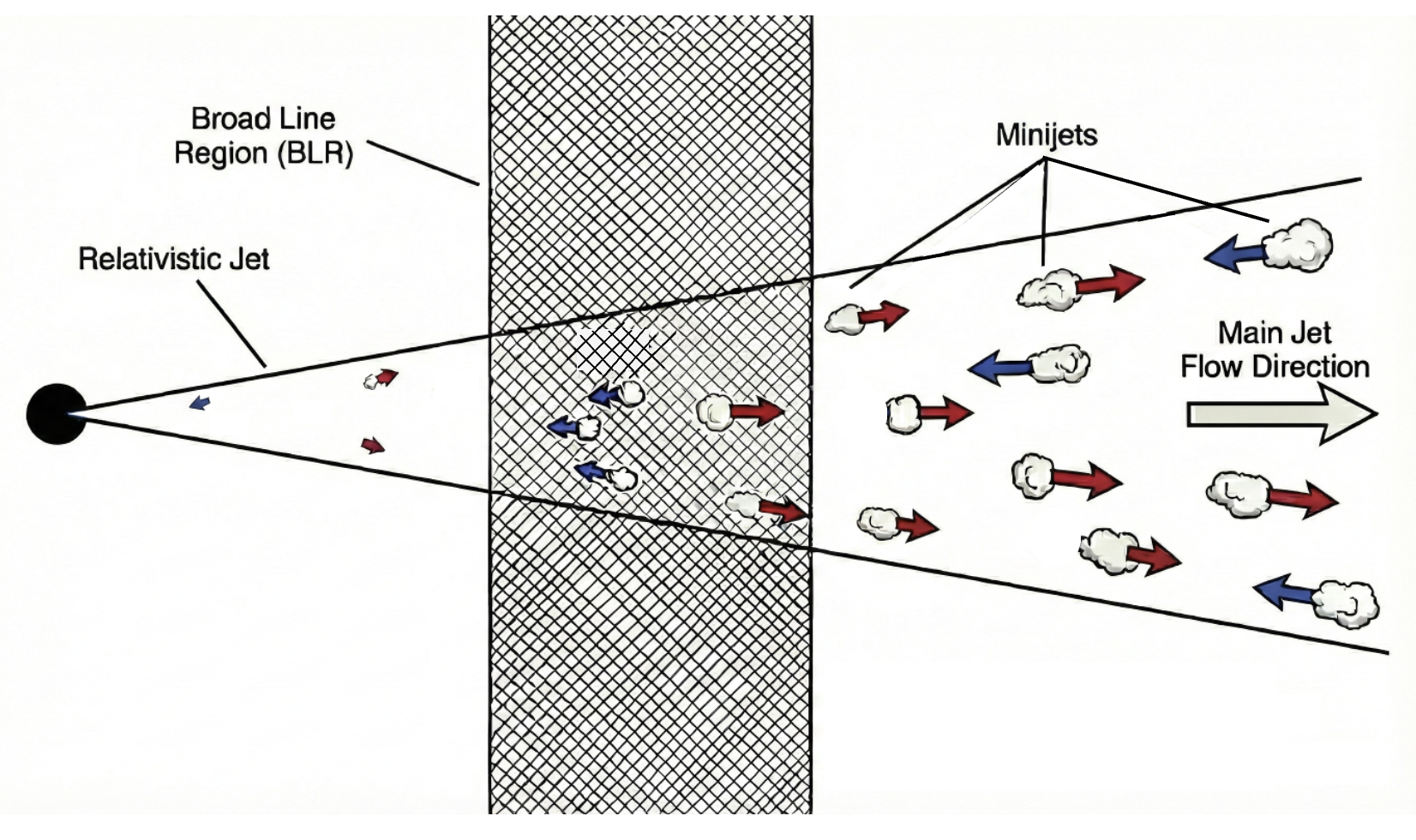}
    \caption{Sketch shows a basic schematic of the minijet model as observed from the lab frame. Red and blue arrows emanating from the minijets denote minijets moving \textit{toward} us and \textit{away} from us, i.e., in and against the direction of bulk jet flow.}
    \label{fig:mini}
\end{figure}

We assume the minijets are inclined at an angle $\theta_{\rm mj}\in[0,\pi]$, which incorporates motion towards and away from us through the Doppler factor $\delta_{\rm mj}$. When these minijets move towards the BLR, the flux is boosted/de-boosted by $\delta_{\rm mj}^{4+2\alpha}$ \citep[e.g.,][]{dermer95} since the GeV emission is primarily external Compton, where $\alpha$ is the GeV spectral index. When they are moving \textit{away} from the BLR along the jet, the beaming pattern follows $\sim\delta_{\rm mj}^{2-2\alpha}$ \citep{dermer10}. Figure \ref{fig:mini} above shows a basic sketch of the idea, that shows minijets both inside and outside the BLR and having a direction either against or along the jet flow.

For the total external Compton GeV flux, one can safely neglect the minijets that are \textit{outside} the BLR, since they are either de-beamed to the lab frame, or the BLR photon density is de-boosted in the minijet frame. Using \cite{dermer95}, the external Compton flux density can be written as

\begin{equation}
\begin{split}
S_{\mathrm{EC}}\left(\epsilon_{\mathrm{obs}},\Omega_{\mathrm{obs}}\right) &= \delta_{\rm mj}^{4+2 \alpha} \frac{c \sigma_{\mathrm{T}} u_{\mathrm{iso}}^* \mathcal{Q}_e}{8 \pi d_L^2 \epsilon^*}(1+z)^{1-\alpha} \\
&\quad \times\left(\frac{1+\cos\theta_{\rm mj}}{1+\beta_{\rm mj}}\right)^{1+\alpha}\left(\frac{\epsilon_{\mathrm{obs}}}{\epsilon^*}\right)^{-\alpha}
\end{split}
\end{equation}

\vspace{10pt}

where $\epsilon^*$ is the typical BLR energy (here taken as $\sim 10$ eV, Lyman-$\alpha$, \citealt{sahakyan22, pod2025}), $u^*_{\rm BLR}$ is the BLR energy density $\simeq 0.02$ erg cm$^{-3}$ for CTA 102 \citep{gasp18}. $\mathcal{Q}_e$ is the electron kinetic luminosity and $d_L$ is the luminosity distance, which is $\sim7000$ Mpc for $z=1.04$.

For this simple model, we start with around $\sim50$ minijets inside the BLR, of the same assumed size. Too low a number of minijets or too high would result in poor statistics or no flaring at all. This number is statistically motivated from \cite{biteau12}, comparing with the types of distributions we have in this work. Relevant literature provide the maximum Doppler factor estimate as $\sim 45$ for this source \citep[e.g.,][]{sahakyan22}. Using this information, we randomly oriented the minijets from $\theta_{\rm mj}\in[0,\pi]$ and chose their Lorentz factors randomly from a power law distribution with index $-2.5$ and minimum Lorentz factor as $2$.

\begin{figure}
    \centering
    \includegraphics[width=\linewidth]{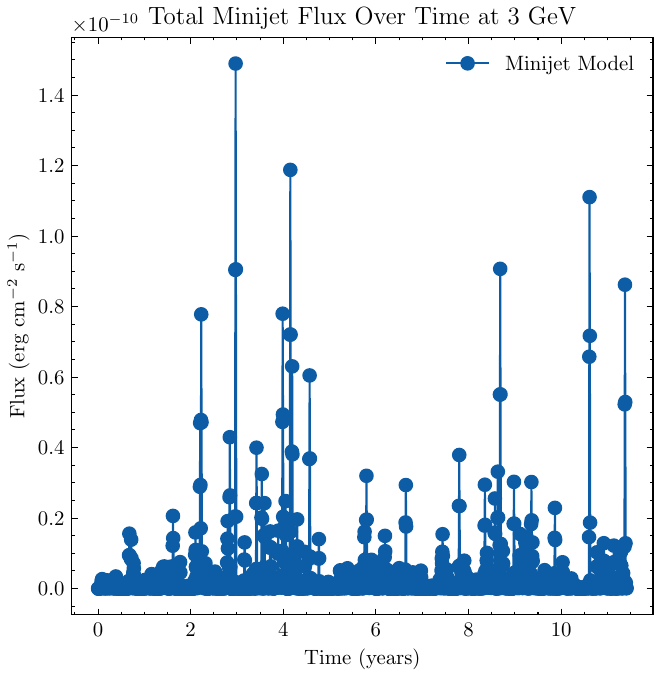}
    \caption{Figure shows a simulated 3 GeV light curve obtained from the superposition of numerous randomly oriented minijets in time, over $\sim$ 10 years.}
    \label{fig:mini_lc}
\end{figure}

Our model assumes that 50 minijets are constantly present, but their viewing angles and Lorentz factors are changing randomly in time. We specifically note here that this differs from the minijet model presented in \cite{biteau12}. They considered an isotropic minijet distribution in the \textit{rest frame of the jet}. In our case, the distribution is isotropic in the \textit{lab frame}, with the jet material being steady and not having a direct link to the minijet beaming. Here we use the jet as only a region where minijets can form. Given that, the constancy of minijets is not unphysical since minijets may be continuously created at a given reconnection rate and may decay at a similar rate. The reconnection rates are such that at all times a fixed average number of minijets is maintained. Further complexities are out of the scope of this paper. For the above parameters, we generated a light curve in arbitrary time units. In order to transform to real time units, one needs to consider a typical reconnection timescale (or equivalently a flare duration), which is $\sim 1$ hour for CTA 102 \citep{geng22}. Figure \ref{fig:mini_lc} shows the result at 3 GeV.

In Figure \ref{fig:mini_lc} one finds multiple flares through a $\sim$ 10 year period. The shape and rate of occurrence of flares is sensitive to the the parameters provided above. The point to note, however, is just that bigger flares are present among a sea of very small flares and they are much less probable than the smaller ones. This is natural since a ``lining up" event is one of very low probability, corroborating the fact that GeV flares are occasional/rare and not continuous. One must not obviously expect that the emitting regions/minijets remain ``same" over time : as internal shocks and magnetic reconnection events create such regions which then emit and dissipate. A series of papers on radio flaring and kinematics in CTA 102 by Christian Fromm \citep{fromm11,fromm13b,fromm13a} suggested possible standing shock - propagating shock interactions as a cause of possible flaring in the radio. Internal colliding shocks have been observed in kilo-parsec scale optical jets \citep{meyer3c264} even. If we assume all the flaring, small or large, was caused by such shock-shock interactions of varying strengths (to reproduce different flare peaks), a change in the skewness of the logarithmic flux distribution is not expected after the extreme flare since in this scenario all flares are fundamentally ``equal", only differing in shock strengths. This is not the case and hence we have discussed alternatives in this section.

\subsection{Magnetic reconnection and a modified log-normal power-law distribution}
\label{sec:mlp}

The change in the flux distribution's skewness can be better explained using magnetic relaxation \citep[e.g.,][]{taylor86, yeates10, zrake17}. The pre-flare light curve had a tangled magnetic field, with no clear global structure. Smaller flares in the pre-flare light curve can be explained by small-scale reconnection events followed by ``lining up". It would not be unwise to believe that a large number of magnetic reconnection (or more generally energy injection) events occurred in the jet prior/during the flare, in addition to ``lining up". The combined effect produced the flare and ``relaxed" the magnetic field to a lower energy and a more ordered configuration. Ordering of the fields would reduce the rate of reconnection events, and thereby the number of minijets. Further, lowering of the magnetic field energy density would drastically reduce the SSC contamination to the GeV. The former effect is much more dominant and a combination of both results in a more ``relaxed" flux dominated state without a pre-flare-like tail. The decrease in the skewness through the flare, or a hint of progression toward 0 skewness support this observation.

\begin{figure}
    \centering
    \includegraphics[width=\linewidth]{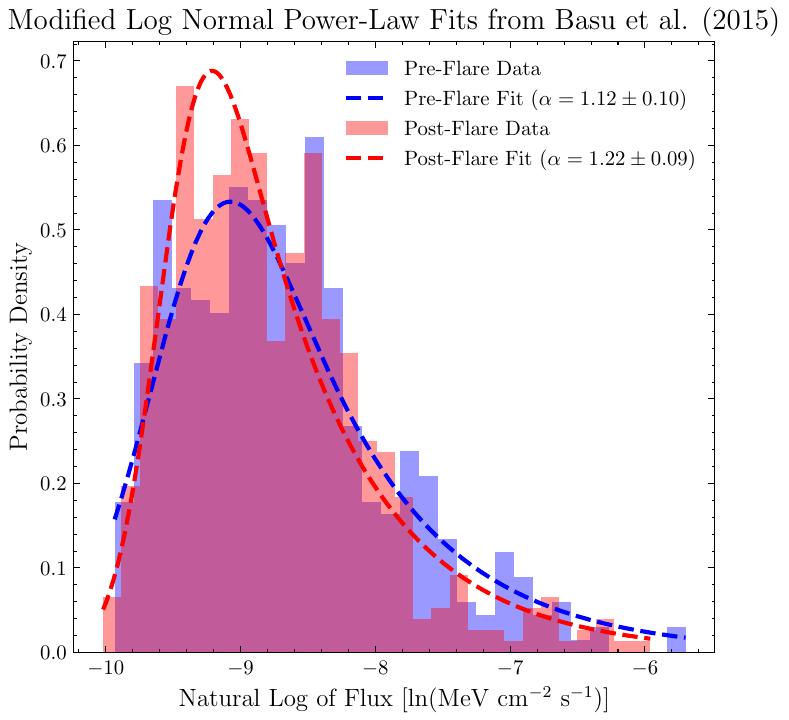}
    \caption{Figure shows the MLP distribution from \citealt{basu2015} fit to the pre and post-flare flux distributions. The blue and red histograms as well as the dotted lines show the histograms and fit for the pre and post-flare fluxes. $\alpha$ is smaller and higher in the pre and post-flare states, in accordance with the expectation of faster and slower electronic injection respectively, as discussed in the text.}
    \label{fig:mlp}
\end{figure}

To better understand the distribution, we used the modified log-normal power-law (MLP) distribution model from \cite{basu2015} for this case. The MLP distribution essentially takes into account the time-evolution of a starting log-normal distribution through a paramater named $\alpha=\Delta/\eta$, where $\eta$ is a typical ``growth" rate and $\Delta$ is a ``decay" or ``stopping" rate for that same growth. It is given as

\begin{equation}
\begin{split}
f(m) &= \frac{\alpha}{2} \exp \left(\alpha \mu_0 + \frac{\alpha^2 \sigma_0^2}{2}\right) m^{-(1+\alpha)} \\
&\quad \times \operatorname{erfc}\left[\frac{1}{\sqrt{2}}\left(\alpha \sigma_0 - \frac{\ln m - \mu_0}{\sigma_0}\right)\right], \quad m \in [0, \infty),
\end{split}
\end{equation}

where \(\alpha=\Delta / \eta\), and $\mu_0$ and $\sigma^2_0$ are the mean and variances respectively. Figure \ref{fig:mlp} shows the fit for both the pre and post-flare flux distributions. We find that $\alpha=1.12\pm0.10$ and $\alpha=1.21\pm0.09$ for the pre and post-flare flux distributions respectively, i.e., they are different at a $\gtrsim 1\sigma$ level. If one interprets $\eta$ as a typical flaring/electron injection rate and $\Delta$ as a particle escape/cooling or a flare decay rate, it implies that the pre-flare blazar had a longer stopping timescale and/or a smaller electron injection timescale. In contrast, the post-flare blazar would be akin to slower injection and a faster decay rate. This is in accordance with our above interpretation where the rate of reconnection events have gone down (drop in $\eta$) and the flux levels have stabilized through the 2017 flare.

\begin{figure}[h!]
    \centering
    \includegraphics[width=\linewidth]{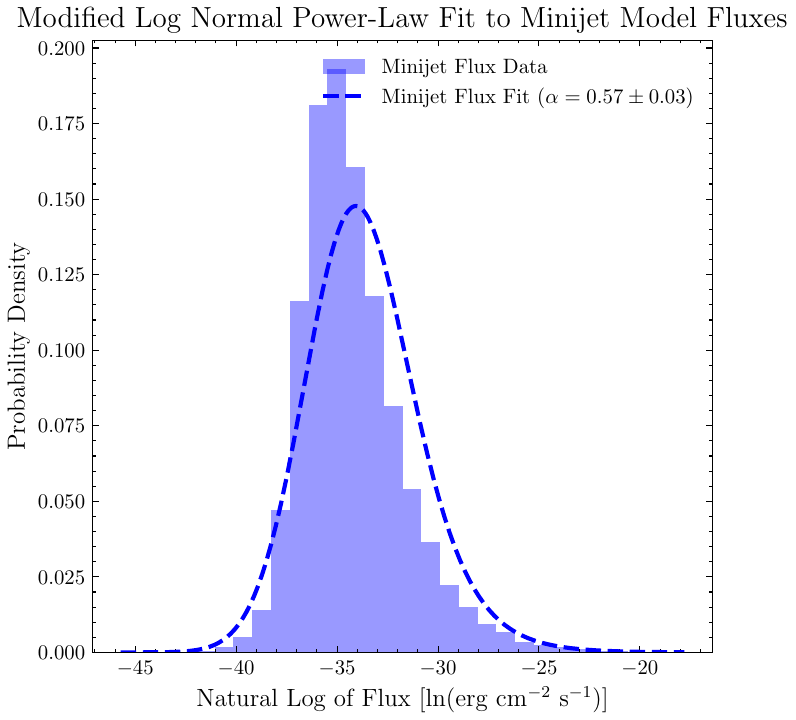}
    \caption{Figure shows the MLP fit to the simulated 3 GeV fluxes from the minijet model. While it is evident that the MLP model captures the long tail of the distribution, the peak is not well predicted.}
    \label{fig:mini_mlp_fit}
\end{figure}

We further attempted to fit the \textit{simulated} minijet flux data that incorporates the random speeds and geometry of a distribution of minijets. Figure \ref{fig:mini_mlp_fit} shows the MLP fit to the flux in Figure \ref{fig:mini}. The longer tail is evident and a usual Gaussian struggles to fit it (skewness $\sim1.1$). The MLP fit captures the tail with an $\alpha=0.57$, while struggling with the peak. While not perfect, this cements the claim that a minijets in a jet model can not only simulate GeV flares, but also possibly capture the nature of their flux distribution, which is \textit{not} log-normal in CTA 102.

\section{Conclusions}

1. Our analysis of the 18 year Fermi-LAT light curve of blazar CTA 102 shows logarithmic flux distribution that is not normal, both before and after the 2017 flare, and instead have a hard tail, deviating from usual expectations from the literature.

2. We find a reduction in skewness of the logarithmic distribution from more positive to less positive values across the flare, where flux from the tails moved ``into" the populated part of the distribution.

3. We explain the observed GeV flaring (both small and large) through a single mechanism. The departure from log-normality can be explained through a modified version of the minijets-in-a-jet model of \cite{biteau12}. We claim and test that the flaring events are only instances of many randomly oriented minijets \citep{biteau12} moving \textit{towards} the BLR and the line of sight, where the rest-frame photon energy density is highly boosted. This is an instance of rare occurrence, tying with the rarity of flares.

4. The pre-flare state possibly had a disordered/tangled magnetic field. The minijets were produced through small-scale magnetic reconnection (energy injection) which then lined up accordingly towards the BLR and the line of sight to produce the smaller flares. The 2017 flare was a result of a large reconnection event, or many happening together, that resulted in a burst of emission. However, this subsequently reduced the magnetic field energy and made it more ordered. A henceforth reduction in magnetic reconnection events and increased synchrotron self-Compton (SSC) contamination resulted in a deletion of the tail seen in the pre-flare histogram, as in Figure \ref{fig:hist}, and a statistically significant decrease in skewness.

5. A modified log-normal power-law distribution (MLP) \citep{basu2015} could be clearly fit to both the pre-flare and post-flare flux distributions. The parameter $\alpha$ defined in the text is higher when the electronic injection rate is slower and vice versa. The fit resulted in an $\alpha$ that is higher in the post-flare state by $\sim1\sigma$ than the pre-flare state. This is in accordance with the idea of magnetic relaxation which would reduce the rate of magnetic reconnection events in the post-flare state. Further, the MLP could also be fit to the simulated minijet flux distribution, cementing the strong relation between minijets and GeV flaring in blazars in general and CTA 102 in particular.

6. A combination of geometric and radiative ideas can fully explain the observed departure from log-normality as well as the transition between a high-tailed unsteady state (pre-flare) to a less-tailed steady state (post-flare).

\section{Acknowledgments}

We acknowledge the support of the Department of Atomic Energy, Government of India, under the project 12-R\&D-TFR-5.02-0700. This work has used Google Gemini 2.5 Pro \citep{gemini2025} for the production of Figure \ref{fig:mini}.

\bibliography{version3}{}
\bibliographystyle{aasjournalv7}



\end{document}